# Luminescent properties of Bi-doped polycrystalline KAlCl$_4$

A.A. Veber[1*], A.N. Romanov[2,3], O.V. Usovich[4], Z.T. Fattakhova[5], E.V. Haula[5], V.N. Korchak[5], L.A. Trusov[4], P.E. Kazin[4], V.B. Sulimov[2,3] and V.B. Tsvetkov[1].

[1]*A.M. Prokhorov General Physics Institute, Russia Academy of Sciences, 38 Vavilov Str., 119991, Moscow, Russia*
[2]*Research Computer Center of M.V. Lomonosov Moscow State University, 1 Leninskie Gory, Build. 4, 119992 Moscow, Russia*
[3]*Dimonta Ltd.,15 Nagornaya Str., Build. 8, 117186 Moscow, Russia*
[4]*Department of chemistry, M.V.Lomonosov Moscow State University, 1 Leninskie Gory, Build.3 119991 Moscow, Russia*
[5]*N.N. Semenov Institute of Chemical Physics, Russian Academy of Sciences, 4 Kosygina Str., 119991 Moscow, Russia*

*Corresponding author. **E-mail**: alexveb@gmail.com, **Tel.**: +7(499)503-8274, **Fax**: +7(499)135-0270

**Abstract**

We observed an intensive near-infrared luminescence in Bi-doped KAlCl$_4$ polycrystalline material. Luminescence dependence on the excitation wavelength and temperature of the sample was studied. Our experimental results allow asserting that the luminescence peaked near 1 μm belongs solely to Bi$^+$ ion which isomorphically substitutes potassium in the crystal. It was also demonstrated that Bi$^+$ luminescence features strongly depend on the local ion surroundings.

**1 Introduction**

Bismuth doped media became of great interest as laser materials in 2001 when the luminescence in range of 1-1.6μm of Bi-doped silicate glass was observed by Fujimoto et al. [1]. Reported luminescence features have spurred the interest in Bi-doped materials as novel prospective active media for lasers and amplifiers for telecommunication, scientific, medical and other purposes. The fact that observed NIR luminescence radically differed from known luminescence of Bi$^{3+}$[2] and Bi$^{2+}$[3] ions initiated the search of the true Bi-related optical centers which is responsible for the phenomena.

To date optical amplification[4,5] and laser operation[5,6] has been demonstrated from 1.1 to 1.5μm. Luminescent properties of multitude Bi-doped glasses (silicate [1], germanate [7], phosphate [8], boro-alumino-phosphate[9], sulfide [10], fluoride [11], chloride [12]), photonic crystals [13], crystals (RbPb$_2$Cl$_5$ [14], BaB$_2$O$_4$ [15], BaF$_2$ [16], Ba$_2$P$_2$O$_7$ [17], CsI [18], zeolite [19], sodalite [20]) as well as Bi-containing crystals (Bi$_5$(AlCl$_4$)$_3$ [21], Bi$_8$(AlCl$_4$)$_2$ [22]) were investigated. Despite many suggested Bi-centers and the continuing controversy, there are more and more evidences that some of low-valence bismuth species like Bi$^+$ ions and subvalent Bi clusters (Bi$_2^{4+}$, Bi$_5^{3+}$, Bi$_8^{2+}$ etc) are responsible for observed NIR luminescence.

Previously it was shown that it is possible to obtain low-valence luminescent Bi species in ZnCl$_2$-AlCl$_3$ glass by controlled Bi$^{3+}$ reduction [12]. Three different luminescent centers were observed in the glass simultaneously: Bi$^+$, Bi$_5^{3+}$ and, probably, Bi$_2^{4+}$. In present work we have attempted to obtain medium with a single luminescent center, namely Bi$^+$. KAlCl$_4$ crystal was selected as a host medium for the investigation. Due to the similarity of K$^+$ ionic radius and Bi$^+$ ionic radius [23] we supposed, that Bi$^+$ can isomorphically substitute potassium in KAlCl$_4$, crystallized from Bi$^+$ containing melt.

**2 Experimental**

The Bi-doped KAlCl$_4$ specimens were synthesized by melting at 270-300°C and subsequent crystallization upon cooling the mixture of dry AlCl$_3$ (99.999%) KCl (99.999%) and BiCl$_3$ (99.998%). The synthesis was performed in evacuated and sealed fused silica cell. All manipulations with the starting chlorides (weighing and transfer to fused silica cell) were performed in argon-filled glovebox Labconco (<2ppm H$_2$O) due to their extreme hygroscopicity. The aluminium chloride was taken in excess relatively to KAlCl$_4$ stoichiometry (39% KCl, 59% AlCl$_3$, 2% BiCl$_3$) to ensure the Lewis acidity of chloride melt. The Lewis acidity of melt shifts the complex equilibrium between Bi metal, BiCl$_3$, subvalent Bi$_5^{3+}$ and Bi$^+$ toward Bi$^+$ formation (by capturing basic Cl$^-$ anions) [24]:

$$2Bi + BiCl_3 \leftrightarrow 3Bi^+ + 3Cl^- \qquad (1)$$

$$Bi_5^{3+} + BiCl_3 \leftrightarrow 6Bi^+ + 3Cl^- \qquad (2)$$

Despite the non-stoichiometric melt composition, the primarily crystallized phase was KAlCl$_4$ [25]. Bismuth monovalent cation in melt was prepared by controlled reduction of BiCl$_3$ by zinc metal, placed in special appendix of fused silica cell:

$$BiCl_3 + Zn \rightarrow Bi^+ + ZnCl_2 + Cl^- \qquad (3)$$

Even in acidic melts, 10-100 fold excess of BiCl$_3$ over subvalent bismuth species is needed to significantly shift the equilibrium in reaction (2) toward Bi$^+$. That is why, only slight reduction according to Eq. (3) is needed (otherwise, sufficient amounts of unnecessary Bi$_5^{3+}$ will be formed). The degree of reduction can be visually inspected: Bi$^+$ solution in melt possesses blue to violet color, whereas the appearance of brown to orange hues signifies the formation of Bi$_5^{3+}$ via excessive reduction. We prefer to cool and crystallize the melt at the point of reduction, when it attains deep blue-violet color.

After crystallization greenish polycrystalline agglutinated powder layer was formed over the wall of the cell. Due to hygroscopicity the sample remained packed during all spectroscopic experiments.

Photoluminescence spectra were measured using an ARC SpectraPro SP-305 monochromator combined with an ARC ID-441C InGaAs.

Emission decay curves were obtained with Hamamatsu G8372-01 InGaAs photodiode coupled with a wideband (100 Hz–5 MHz) low-noise preamplifier mounted on the exit slit of the monochromator. Q-switched frequency doubled Nd:YAG laser (532nm) or high-power light emitting diode (590nm, 20mW in CW, 30 nm FWHM) were used as excitation sources for the lifetime measurements. LED was electrically modulated by rectangle pulse with width of 400μs and tripled feed current in comparison with CW mode. Trailing edge of the excitation light pulse was about 3 μs in this case.

Osram HBO 150 W Xe-lamp combined with LOMO MDR-12 monochromator was used as variable wavelength light source to obtain luminescence excitation spectra. The measurements were performed at 295 K and 77 K temperatures. All the obtained spectra were corrected for the spectral response of the system.

## 3 Results

At room temperature intense NIR luminescence peaked at 972 nm with FWHM of 130 nm was observed. After cooling to 77K the luminescence band was shifted to 900 nm and became narrower with FWHM of 70nm. No any fine structures/additional peaks were detected with up to 1nm spectral resolution (Fig. 1).

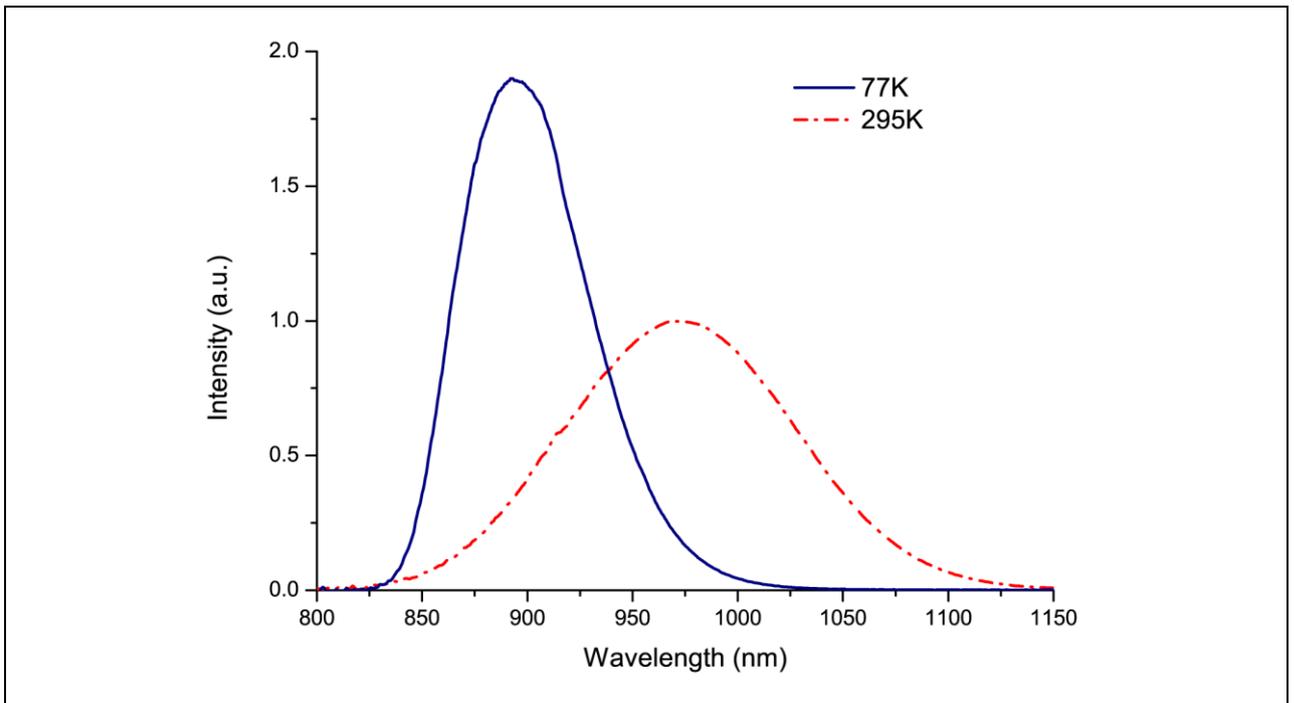

Figure 1. Luminescence spectra measured at 77 and 295 K; excitation light wavelengths were 610 and 600 nm respectively.

Additional investigation of the luminescence band shape by changing the excitation wavelength in the range of 360-930 nm (295 K) and 360-860 nm (77 K) with 10 nm step were carried out. Spectral resolution of exciting channel monochromator was about 7 nm in these experiments. It was revealed that the shape at both temperatures didn't depend on the excitation light wavelength, only the luminescence intensity was changing. Experimental data resulted in complete excitation spectra (Fig.2) for the luminescence centered near 900 nm (77 K) and 975 nm (295 K). Excitation spectra in the range of 550-740 nm were measured additionally with a higher spectral resolution (2 nm).

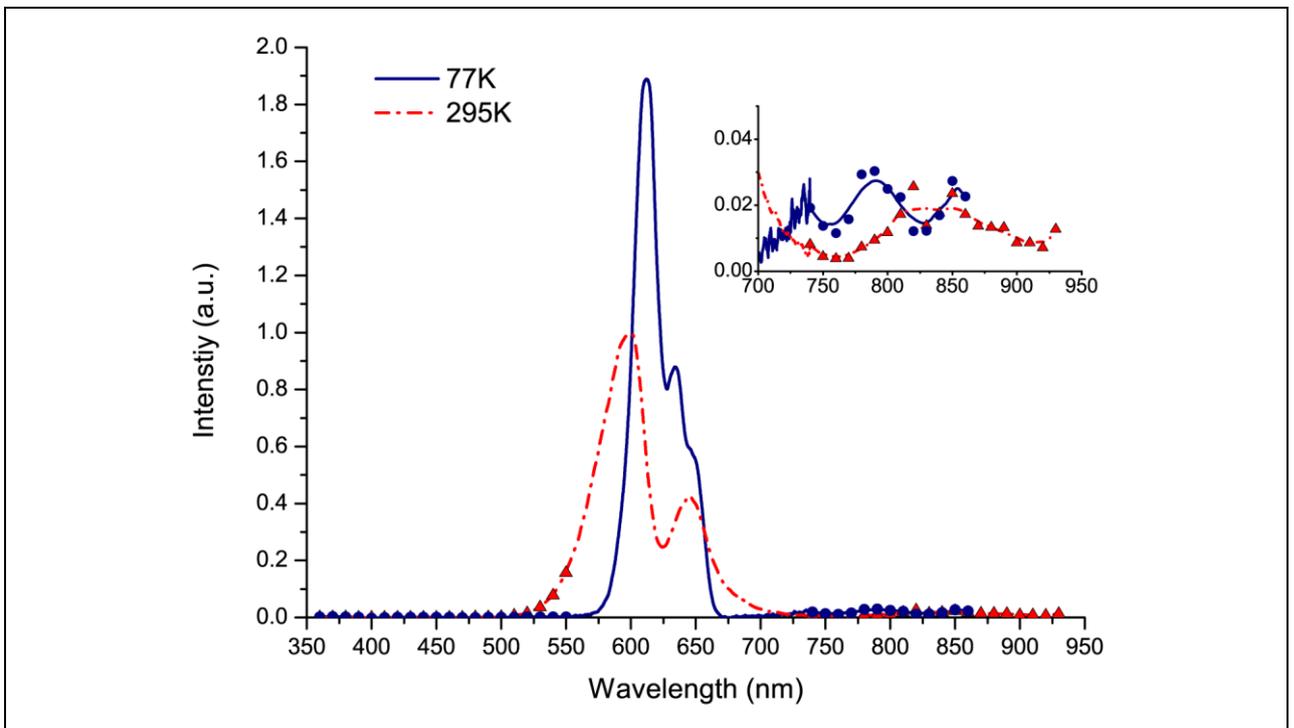

Figure 2. Luminescence excitation spectra measured at 77 and 295 K. Enlarged spectra for 700-950 nm are shown on the inset. The spectra were measured continuously for the range 550-740 nm (2 nm spectral resolution) and step by step (10 nm step, 7 nm spectral resolution) for the 360-550 nm and behind 740 nm.

Two well resolved intense peaks centered at 600 nm (I) and 645nm (II) and weak wide band in the range of 800-900 nm (III) were observed in the excitation spectrum at room temperature. At 77K all the excitation peaks became narrower and an additional structure of the excitation band II was observed. It consisted of a two peaks centered at 635 nm (II.a) and 650nm (II.b). Bands I and III were shifted to 610 nm and 720-820 nm respectively. True shape of the excitation band III may differ from obtained one due to low luminescence level at this range. No luminescence was observed for excitation wavelengths less than 460 nm and 560 nm for both 295 and 77 K measurement temperatures.

Luminescence lifetime measurements at room temperature initially were performed with using of second harmonic of Q-switched Nd:YAG laser. The single exponential decay curve was observed with a lifetime of 525 μs at room temperature. Unfortunately at 77 K the excitation spectrum is cut off at longer wavelength (Fig.2) and no luminescence was observed even with 60 MW/cm$^2$ pulse intensity. Therefore amber LED was used for further lifetime measurements in the range 77-295 K (Fig.3).

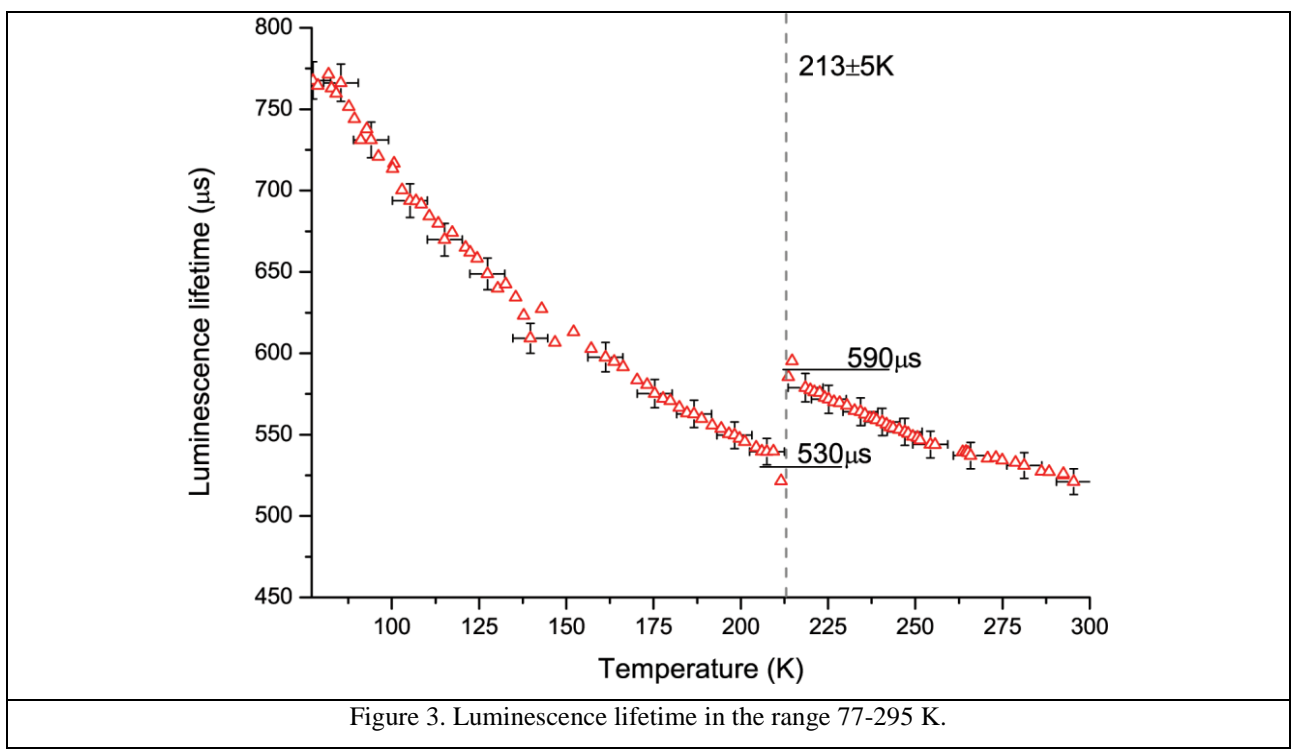

Figure 3. Luminescence lifetime in the range 77-295 K.

Luminescence intensity was monitored at 940 nm. At 77 K the value of lifetime was 765 μs. With increasing temperature the lifetime remained constant up to 85 K then monotonically decreased up to 212 K. Further heating resulted to abrupt lifetime growth in 60 μs at 214 K. Then lifetime again is monotonically decreased up to room temperature. Lifetime values at 295 K were equal being obtained at different excitation wavelength.

**4 Discussion**

The single peak with the shape independent on excitation wavelength was observed in luminescence spectra indicating that it belongs to the single luminescence center. Furthermore the increase of intensity and simultaneous narrowing of the luminescent peak while cooling the sample shows that the dopant enters into the lattice as a substitution of the lattice ion. For this case the ionic radii of K$^+$ and Bi$^+$ being 132 pm and 148 pm [23] respectively, are well matched. Bi$^{3+}$ ion is significantly greater than Al$^{3+}$ (102 pm vs 52 pm [23]) and can hardly occupy this site. In addition taking into account the fact that reduction reaction was carried out during the synthesis one can conclude that Bi$^+$ is the most preferable bismuth valence form which can exist in KAlCl$_4$ crystal. No charge compensation is needed in this case also. It is hardly to assume that some subvalent bismuth cluster (Bi$_2^{4+}$, Bi$_5^{3+}$ etc) can be embedded into the crystal.

Energies of observed excitation and luminescence bands coincide well with experimentally and theoretically obtained Bi$^+$ energy level system (see Table 1).

*Table 1. Excited states energies of Bi$^+$ ion in different media*

| Excited state | Energy, 10$^3$cm$^{-1}$/ Wavelength, nm | | | |
|---|---|---|---|---|
| | NaCl-AlCl$_3$ [26] measured | KCl [27] calculated | KAlCl$_4$ (this work) measured | |
| | | | 295K | 77K |
| $^3$P$_2$ | 17.1/585 | - | 16.67/600 | 16.39/610 |
| | 15.2/658 | 15.45/647 | 15.5/645 | 15.75/635 |
| | 14.4/694 | 15.05/665 | - | 15.38/650 |
| $^3$P$_1$ | 12.7/787 | - | 11.1-12.5/800-900 | 12.2-13.9/720-820 |
| | 11.1/900 | 11.42/875 | 10.25/976 | 11.1/900 |

Obtained experimental results is very similar to reported before for RbPb$_2$Cl$_5$ [14] and to some extent for Ba$_2$P$_2$O$_7$ [17] and BaF$_2$ [16] Bi-doped crystals. In these crystals the rubidium and barium effective ionic radii (r$_{Rb}^+$= 148 pm, r$_{Ba}^{2+}$= 137 pm [23]) also permit their isomorphous substitution by Bi$^+$, although in the latter case the charge compensation is needed. The different charge compensating surrounding of Bi$^+$ in barium crystals produce the more complicated excitation and emission spectra, in compare with Bi-doped KAlCl$_4$ and RbPb$_2$Cl$_5$.

Thus we can conclude that observed luminescence definitely belongs to Bi$^+$ ion in potassium position of the crystal.

Luminescence lifetime, excepting observed jump at 213±5K, is monotonically decreasing while increasing the temperature. During consideration of the crystal structure it was found out that KAlCl$_4$ crystal is thermally trimorphic [28] and can exists in three different crystal phases: low- (I), room- (II) and high- (III) temperature ones. The previously determined transition points between phases I↔II (the 1$^{st}$ phase transition) and II↔III (the 2$^{nd}$ phase transition) are 196 K and 387 K respectively [28]. Since NIR luminescence of Bi$^+$ ion corresponds to the transition to the $^3$P$_1$ → $^3$P$_0$ transition in an open external 6p shell (which is rather diffuse) it should strongly depend on the local structure of the ion surroundings. Thus the observed abrupt lifetime changing can be explained by occurring phase transition and as a result of changing in the Bi$^+$ local surrounding. The deviation of abrupt lifetime changing temperature, observed here, from previously reported KAlCl$_4$ phase transition point[28] can be related with Bi-doping of the crystal and/or considerably higher heating speed in our experiments. Bismuth in this case serves as a phase transition sensor. The changing of impurity luminescence properties due to phase transition of the host media was previously observed for Sm$^{2+}$ [29], Cr$^{3+}$ [30], Mn$^{2+}$ [31], Pr$^{3+}$ [32] and other ions in different crystal hosts. To our knowledge this work is the first evidence of such effect in Bi-doped luminescent media.

Probably some other luminescence features also undergo abrupt changes due to phase transition. Thus at present it is hard to determine what exactly has a greater influence on the luminescence properties: observed phase transition or cooling, therefore additional studies is needed. Currently we are actively investigating all the features of Bi-doped KAlCl$_4$ crystal and results of the research will be published later.

**5 Conclusions**

Bi-doped KAlCl$_4$ chloride crystal was prepared under the conditions preferable for Bi$^+$ formation. Investigation of NIR luminescence properties of Bi-doped KAlCl$_4$ crystal and experimental data analysis allow to conclude that Bi$^+$ ion solely is responsible for observed NIR luminescence as an optical center substituting the potassium ion in the crystal lattice. The observed abrupt luminescence lifetime change is due to phase transition occurred in the host crystal demonstrating the strong dependence of Bi$^+$ optical properties on local surrounding of the ion.

## Acknowledgments

Financial support from the contract № 07.514.11.4059 of October 12th 2011 between Russian Ministry of Education and Science and "Dimonta" Ltd. is gratefully acknowledged.